\documentclass[conference]{IEEEtran}

\usepackage{booktabs}
\usepackage{graphicx}
\usepackage{tabularx}
\usepackage{amsmath}
\usepackage{amssymb}
\usepackage{textcomp}
\usepackage{enumitem}
\usepackage{booktabs}
\usepackage{multirow}
\usepackage{siunitx}
\usepackage{balance}
\usepackage{xcolor}
\usepackage{cite}
\usepackage[para,online,flushleft]{threeparttable}
\usepackage[linesnumbered, noend, algoruled,boxed,lined]{algorithm2e}

\usepackage{geometry}
\ifCLASSINFOpdf
\else
\fi

\begin{document}
\title{\Large \bf HL-Pow: A Learning-Based Power Modeling Framework for High-Level Synthesis}

\author{\IEEEauthorblockN{Zhe Lin\IEEEauthorrefmark{1}, Jieru Zhao\IEEEauthorrefmark{1},  Sharad Sinha\IEEEauthorrefmark{2} and Wei Zhang\IEEEauthorrefmark{1}}
	\IEEEauthorblockA{\IEEEauthorrefmark{1}Hong Kong University of Science and Technology, Hong Kong}
	\IEEEauthorblockA{\IEEEauthorrefmark{2}Indian Institute of Technology (IIT) Goa, India\\
		\{zlinaf, jzhaoao\}@connect.ust.hk,  sharad\_sinha@ieee.org, wei.zhang@ust.hk}}
\maketitle


\setlength{\abovedisplayskip}{3pt}
\setlength{\belowdisplayskip}{3pt}

\begin{abstract}
High-level synthesis (HLS) enables designers to customize hardware designs efficiently. However, it is still challenging to foresee the correlation between power consumption and HLS-based applications at an early design stage. To overcome this problem, we introduce HL-Pow, a power modeling framework for FPGA HLS based on state-of-the-art machine learning techniques. HL-Pow incorporates an automated feature construction flow to efficiently identify and extract features that exert a major influence on power consumption, simply based upon HLS results, and a modeling flow that can build an accurate and generic power model applicable to a variety of designs with HLS. By using HL-Pow, the power evaluation process for FPGA designs can be significantly expedited because the power inference of HL-Pow is established on HLS instead of the time-consuming register-transfer level (RTL) implementation flow. Experimental results demonstrate that HL-Pow can achieve accurate power modeling that is only 4.67\% (24.02 mW) away from onboard power measurement. To further facilitate power-oriented optimizations, we describe a novel design space exploration (DSE) algorithm built on top of HL-Pow to trade off between latency and power consumption. This algorithm can reach a close approximation of the real Pareto frontier while only requiring running HLS flow for 20\% of design points in the entire design space.
\end{abstract}
\IEEEpeerreviewmaketitle

\section{Introduction}
High-level synthesis (HLS)~\cite{cous09} automates the process of translating applications described by high-level languages (e.g., C++ and Python) into register-transfer level (RTL) designs. With the aid of HLS tools, designers targeting hardware implementation for field-programmable gate arrays (FPGAs) or application-specific integrated circuits (ASICs) are no longer required to dig deep into low-level hardware details, such as the micro-architectures of individual components and the interconnection between them. Besides this, modern HLS tools have the capability to give relatively good estimation of performance and resource utilization for the created hardware, and also deliver a series of design knobs, or so-called directives, to help designers tune the two aforementioned design metrics. As a result, the productivity and flexibility brought by HLS notably speed up the development process of hardware designs, and also open up an opportunity for efficient design space exploration (DSE)~\cite{liu13, fer18, gzhong17, meng16, dong16, lfer18}. However, off-the-shelf HLS tools~\cite{vivadohls} are still lacking in mature power analysis techniques, making it difficult to clearly observe the influence of different optimization strategies of HLS on power consumption.

Power consumption is a primal concern for many hardware designs, especially for portable electronic devices and embedded systems. The common practice to obtain power consumption is through power measurement or estimation, both of which require designers to spend substantial effort. First, RTL designs are created by designers either manually or through HLS. Afterwards, the RTL implementation flow, including logic synthesis, placement and routing, is applied to the provided RTL designs for the generation of gate-level details. For power measurement, the designs are implemented on real systems, and power consumption can be measured onboard by monitoring devices. For power estimation, gate-level simulation is performed with real input vectors to capture switching activities of the IO and internal signals. Thereafter, a prebuilt analytical power model~\cite{liu94} provided by the design tool is applied to compute power consumption given the gate-level details and signal activities. After obtaining power values, designers can accordingly refine the hardware architectures in pursuit of higher performance or power efficiency, and run the above design flow again for verification. In general, the creation of power-efficient designs usually necessitates multiple iterations of power evaluation and design refinement, which results in a long design time and low productivity.

Some state-of-the-art works~\cite{lee15, zhe18, zuo15, liang18} have presented power modeling techniques to accelerate the power analysis process for hardware designs; however, each of these methods exhibits some of the following drawbacks: 1) each of the power models generated by these methods is customized for an individual design and not applicable to others, 2) their modeling process for each target design requires multiple rounds of power characterization following the slow RTL implementation flow, and 3) it is difficult to migrate their techniques to new platforms due to their dependence on specific hardware modeling expertise. Putting it all together, designers must familiarize themselves with the modeling steps and make great effort to build a specialized power model for every target design, thus incurring high labor intensity.

In light of the above considerations, in this work, we investigate advanced modeling techniques to provide power prediction for FPGA designs at an early design stage, and also strive to speed up power-oriented exploration of hardware designs. Specifically, we propose HL-Pow, a learning-based power modeling framework for HLS designs. Our modeling framework features wide applicability and high efficiency compared with state-of-the-art works~\cite{lee15, zhe18, zuo15, liang18}. First of all, HL-Pow offers a modeling strategy with high generalization ability so that various designs can use one well developed model for power prediction without the need of model reconstruction when targeting the same FPGA platform. Second, our methodology can be easily migrated to new platforms without knowing low-level hardware details such as the technology, hardware primitives or macros. Third, the power prediction of HL-Pow for new designs is fast in runtime, as it dispenses with the need to perform the time-consuming RTL-based power estimation or measurement flow. With HL-Pow, DSE can be quickly conducted to investigate the design tradeoff between power and other design metrics provided by HLS. In summary, we demonstrates the following contributions in this work:
\begin{itemize}[noitemsep]
	\item We introduce an automated feature construction flow for rapid identification and extraction of features closely related to power consumption, simply using results generated by the HLS design flow.
	\item We propose HL-Pow, a learning-based power modeling methodology with the ability to achieve accurate, fast and early-stage power estimation for HLS designs, by building the power model only once.
	\item We describe a novel DSE algorithm established on HL-Pow to demonstrate how the tradeoff between latency and power consumption can be effectively and efficiently evaluated by design space sampling.
\end{itemize}

\section{Related Work}
\subsection{Hardware Power Modeling}
\label{ssc: pwr_model}
Studies about hardware power modeling have been conducted at two abstraction levels: low abstraction and high abstraction. Low-level abstraction methods~\cite{bogl00, shao14, dm07, hao15} look into the power consumption of primitive components, and derive overall power consumption by aggregating power of all used primitive components. For this purpose, a library is built in advance for real-time power reference of primitive components. A power characterization process should be conducted to construct a power look-up table, or a so-called macro-model, for each basic component, such as the adder and multiplier. Except that a rich body of basic components should be characterized individually, this power characterization stage should also take into account various use cases, such as signal activity levels, bitwidths and even cell selection variances, thus leading to a large evaluation space to walk through all different situations per component. The large characterization space for all components requires a tremendous amount of development time. What's more, different technologies or standard cell libraries would have their specific design methodologies that are not shared among the others. Based on this, creating this library also depends on developers having a good understanding of all primitive components. 

In contrast, high-level abstraction methods~\cite{lee15, zhe18, zuo15, liang18} view a design as a whole and build an analytical or learning-based model specific to it, which avoids going deep into most low-level hardware details. The works~\cite{lee15} and~\cite{zhe18} are for post-RTL power modeling, while the works~\cite{zuo15} and~\cite{liang18} focus on pre-RTL power modeling and they are close to our work. The work~\cite{zuo15} specifically looks into affine functions, identifies the basic code segment as a tile from the programs, and deduces overall power consumption by summing up power consumption of all tiles. For each application, the tile structure is unique. As a result, given a new application, a tile-based power characterization stage still needs to be carried out through gate-level power simulation. Nevertheless, the power characterization time can be significantly expedited compared with low-level abstraction methods, because only the tile structures instead of a pool of primitive components should be characterized. Another work FlexCL~\cite{liang18} targets OpenCL-to-FPGA design flow. Based on the fact that OpenCL applications tend to show regular behaviors in phases, FlexCL decomposes the execution timeline of a kernel into work-groups, and then further divides work-groups into work-items. The dynamic power model is generated according to these two phase levels. Similar to the work~\cite{zuo15}, FlexCL also involves the fine-grained power characterization for different phases in work-groups and work-items, but the overall characterization overhead is also remarkably reduced compared with low-level modeling techniques. 

The high-level abstraction modeling methods show significant speedup in model creation compared with low-level abstraction methods. However, existing high-level abstraction methods still entail model regeneration for new designs, rely on slow and repetitive power estimation/measurement for power characterization, and can not be easily migrated to new platforms because some critical steps, such as power profiling for particular components or code structures, involve hardware design expertise. To the best of our knowledge, our work for HLS-based power modeling, HL-Pow, is the first work that overcomes all these aforementioned limitations, and finally presents an HLS power modeling framework that can deliver high accuracy, efficiency and generalization ability.

\subsection{Design Space Exploration}
A rich body of research studies DSE for HLS. One direction of automatic DSE is to establish predictive models offline and use brute-force search to retrieve an approximate Pareto frontier between two or more target metrics. The works~\cite{zuo15} and~\cite{liang18} elaborated in Section~\ref{ssc: pwr_model} also provide exhaustive DSE after the power model is developed for an application. Another instance is the MPSeeker~\cite{gzhong17} which evaluates the tradeoff between performance and area by producing a predictive model for early estimation of HLS results and then traversing the complete design space to find optimal points.

An alternative to these methods is to select a subset of design points to feed into HLS and search new design points for exploration according to present HLS results. Due to the difficulties of getting information of all design points in advance, methods developed in this way first selects a small subset of samples as promising candidates to put into HLS execution. After obtaining the results from current sample points, knowledge can be learned and used to navigate the search space for evaluating new candidate points. The knowledge generalization techniques include heuristic methods~\cite{fer18, lfer18} that are specific to their target problems, learning-based methods~\cite{liu13, meng16} to generate predictive models for HLS results, and a combination of them~\cite{dong16} which applies heuristic algorithms and machine learning methods in different stages. 
In our work, we first develop a generic model for rapid power inference of HLS designs, and based on that we present a novel heuristic algorithm to further speed up the DSE to evaluate the latency-power tradeoff by online design space sampling. These two stages are complementary to each other for fast design-time hardware power optimization.

\section{Power Modeling Framework}

\begin{figure}[t]
	\begin{center}
		\includegraphics[width=\linewidth]{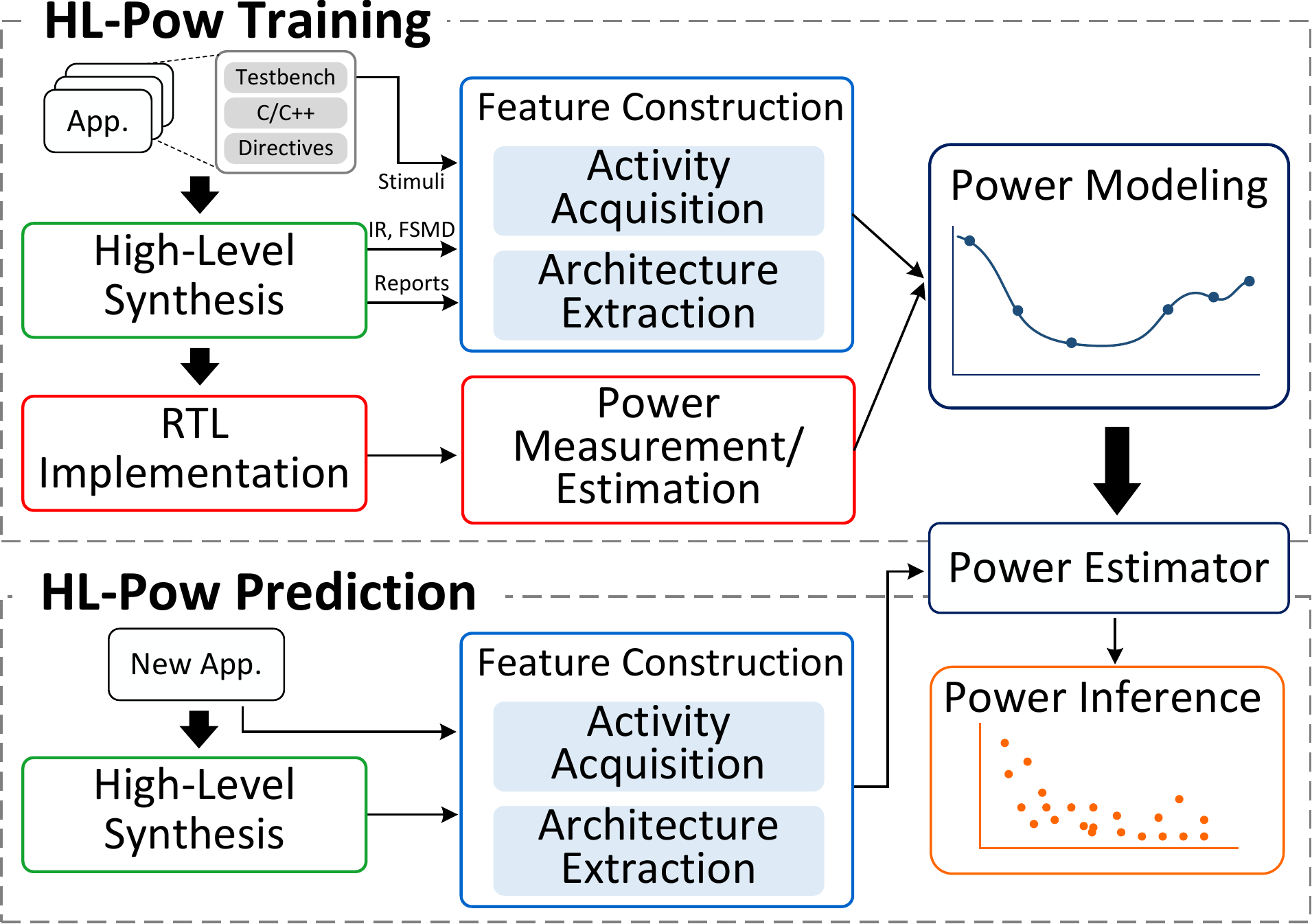}
		\vspace{-6mm}
		\caption{Overview of HL-Pow design flow.}
		\label{fig: flow}
		\vspace{-8mm}
	\end{center}
\end{figure}

Starting with a new platform, the HL-Pow design flow has two phases: 1) power model training with a collection of applications and 2) power inference for new applications. The complete design flow of the HL-Pow framework is depicted in Fig.~\ref{fig: flow}. In the training phase, a number of representative applications described in C or C++ are used to generate training samples for power modeling. Each application is associated with a set of optimization strategies (i.e., directives) to produce a number of design points varying in performance, resource utilization and power consumption. The directives used in this paper are \emph{array partitioning}, \emph{loop unrolling} and \emph{loop pipelining}. The collected design points first pass through the traditional HLS design flow to be converted into synthesizable RTL designs. After that, two major steps are conducted for training sample generation: \emph{feature construction} and \emph{power collection}. For feature construction, we make use of input stimuli, the generated reports and intermediate results from HLS runs to construct features that are of great importance to power consumption. For power collection, the power consumption obtained from estimation or from onboard measurement can be used as ground truth power values, both of which require the design points from HLS to go through the RTL implementation flow. Putting it all together, the feature set and the corresponding true power consumption of each design point constitute a training sample. A training set with multiple samples from different applications is used to build a learning model that maps from features to power consumption.

In the power inference phase, HL-Pow can achieve fast and accurate power prediction for new applications using the well trained power model. Firstly, the new applications, together with the directive configurations to evaluate, are required to go through the HLS design flow. Note that in this stage, RTL code generation can be skipped to save time if the target HLS tool supports the separate execution of different steps in the back-end process. Secondly, the same feature construction step as in the training phase is executed to capture features for new design points. Finally, the created feature set is fed into the prebuilt model for power inference. In this stage, all the steps are solely based on HLS and thus there is no need to invoke the tedious RTL implementation flow along with power estimation or measurement for any design point.

There are two main types of features to acquire: \emph{architecture features} and \emph{activity features}. Architecture features describe the overall design information estimated by HLS tools, while activity features correspond to the switching activities of different hardware components in the target designs. 

\subsection{Data Collection}
\label{ssec: datacollect}
Starting from the HLS front-end execution, the C/C++ source code is first translated into intermediate representation (IR). Some optimizations are also performed by vendor tools at this IR level, such as bitwidth reduction and loop unrolling. With the IR code, the HLS back-end process then conducts control and data flow graph (CDFG) generation, followed by resource allocation, scheduling and functional unit binding. At this stage, the hardware architecture is determined and described by a finite state machine with datapath (FSMD) model. Finally, code generation is executed to convert the generated FSMD model into synthesizable RTL code. 

Using Vivado HLS~\cite{vivadohls} as the design tool for demonstration, some of the data and intermediate results from HLS runs are collected for feature construction: 1) the HLS report (app\_name.verbose.rpt.xml) containing details of the overall design, as described in Section~\ref{ssec: arch}, 2) the IR code (a.o.3.bc) and IR operator information (app\_name.adb), including each IR operator's ID, opcode, type and netlist name corresponding to a hardware component (denoted as RTL operator), and 3) the FSMD model (app\_name.adb.xml) that describes the FSM stages, dataflow, and RTL operator information, including each RTL operator's ID, operand bitwidths and related IR instructions. We identify four types of IR operators that contribute the most to power consumption and can be mapped to RTL operators through ID matching: arithmetic, logic, memory and arbitration operators, as shown in Table~\ref{table: optype}. The activity features introduced in Section~\ref{ssec: act} only account for these operators. Besides the hardware micro-architectures, the operators' switching activities also depend on the input stimuli, which can be collected from real scenarios or generated at random.

\begin{table}[t]
	\centering
	\caption{Operator types and IR opcodes for activity tracking.}
	\vspace{-2mm}
	\label{table: optype}
	\begin{tabular}[width=\linewidth]{c|c}
		\toprule
		\multicolumn{1}{c|}{\textbf{Operator type}} & \multicolumn{1}{c}{\textbf{IR opcode}}\\
		\midrule
		\multicolumn{1}{c|}{Arithmetic} & add, sub, mul, div, sqrt, fadd, fsub, fmul, fdiv, fsqrt\\
		\multicolumn{1}{c|}{Logic} & and, or, xor, icmp, fcmp \\
		\multicolumn{1}{c|}{Memory} & store, load, read, write\\
		\multicolumn{1}{c|}{Arbitration} & mux, select \\
		\bottomrule
	\end{tabular}
	\vspace{-6mm}
\end{table}

\vspace{-1mm}
\subsection{Architecture Features}
\label{ssec: arch}
The power consumption is associated with the scale and complexity of the hardware design and the operating frequency. Therefore, we construct the following architecture features for each design point from the HLS report: 1) FPGA resource utilization estimated by HLS, including look-up table (LUT), flip-flop (FF), digital signal processing unit (DSP) and block random access memory (BRAM); 2) performance, including achieved clock period in nanoseconds and latency in cycles; and 3) the scaling factors (SFs) of the above metrics for the current design to those of the baseline design, respectively, which can be computed as
\begin{equation}
	\label{eq: sf}
	SF_M = \frac{M_{current}}{M_{base}},
\end{equation}
where $M$ represents one of the metrics (i.e., different types of resources, clock period or latency) of the current design, $current$, or the baseline design, $base$, in which no directives are used. In general, the SF is a type of important reference that helps to normalize the resource utilization and performance across different applications. We constructs 11 architecture features in total.

\subsection{Activity Features}
\label{ssec: act}
\begin{figure}[t]
	\begin{center}
		\includegraphics[width=0.95\linewidth]{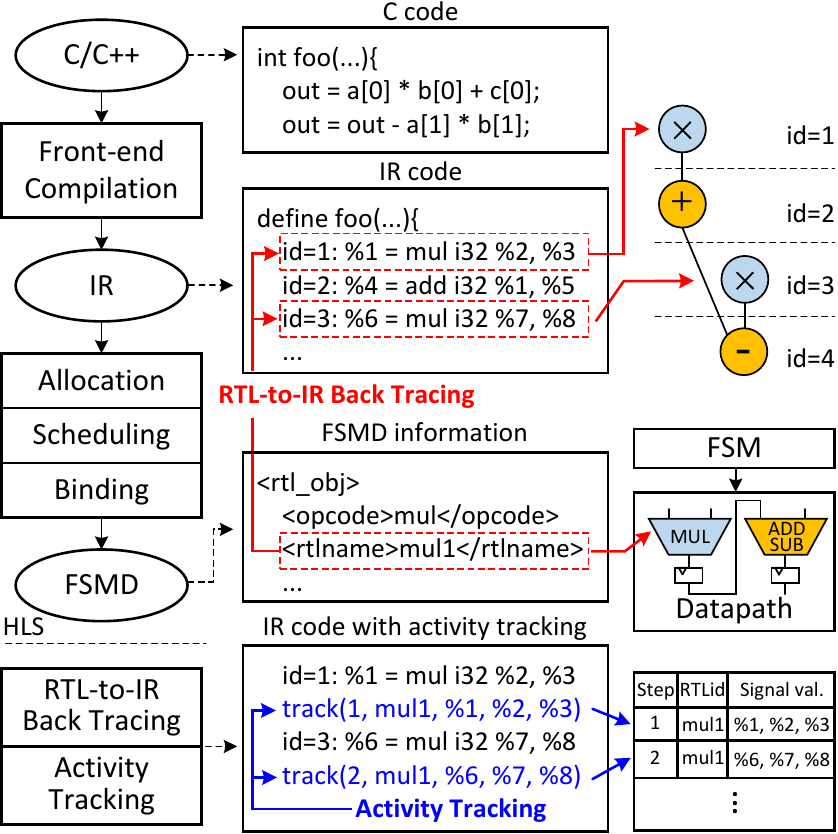}
		\vspace{-2mm}
		\caption{The IR annotator with RTL-to-IR back tracing and activity tracking.}
		\label{fig: actflow}
		\vspace{-9mm}
	\end{center}
\end{figure}

Dynamic power is introduced by signal transitions which dissipate power by repeatedly charging and discharging the load capacitors. Eq.~\ref{eq:dyn} formulates dynamic power $P_{dyn}$ as
\begin{equation}
	\label{eq:dyn}
	P_{dyn}=\sum_{i \in I}\alpha_iC_iV_{dd}^2f,
\end{equation}
which is a function of signal switching activity $\alpha_i$, capacitance $C_i$ on the net $i$, supply voltage $V_{dd}$ and operating frequency $f$. It is conceivable that switching activities of different RTL operators are critical indicators for dynamic power consumption. In HL-Pow, an automatic design flow is introduced to capture the switching activities of different components, and construct activity features using them. To reduce runtime overhead, the design flow targets IR-level activity extraction, instead of the time-consuming RTL-based simulation. The HLS intermediate results elaborated in~\ref{ssec: datacollect} (a.o.3.bc, app\_name.adb, and app\_name.adb.xml) are used during the construction of activity features. Finally, an IR annotator, an activity generator and a histogram constructor are incorporated in this design flow. 

\textbf{IR Annotator.} The IR annotator instruments RTL operators with functions to keep track of their switching activities. The two main steps in the IR annotator are \emph{RTL-to-IR back tracing} and \emph{activity tracking}, as shown in Fig.~\ref{fig: actflow}. The RTL-to-IR back tracing is based on the observation that multiple IR operators can be mapped to the same RTL operator due to scheduling and resource sharing in the HLS back-end process, as depicted in the right-hand side of Fig.~\ref{fig: actflow}. Therefore, multiple IR operations may contribute to the activities of one RTL operator in different time steps. In the IR code, we trace back the RTL operators to their corresponding IR operators with the opcodes shown in Table~\ref{table: optype}. This is done by matching the netlist name between IR operators and RTL operators in the FSMD model. Following the RTL-to-IR back tracing process, we instrument the IR code with an activity tracking function after each IR operator to record the values of input and output signals and the associated RTL operator ID of this IR operator. After all the above steps, an annotated IR is generated. This IR annotator is developed within the LLVM compiler toolchain~\cite{llvm04}.

\begin{figure}[t]
	\begin{center}
		\includegraphics[width=0.9\linewidth]{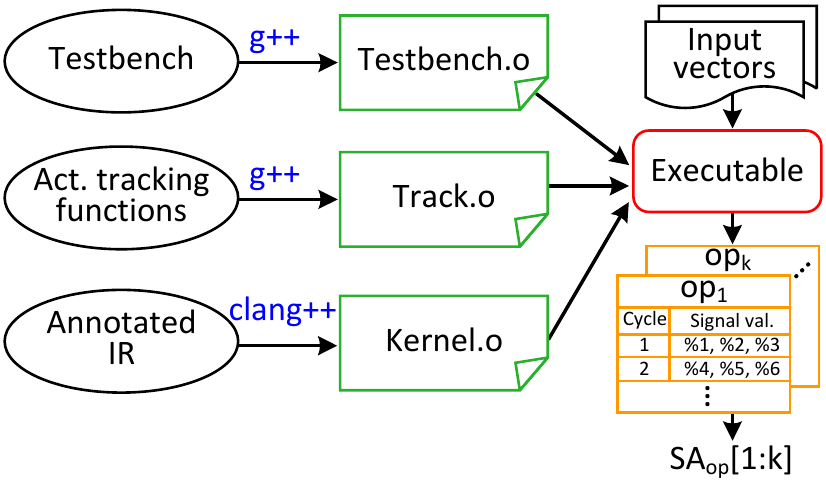}
		\vspace{-4mm}
		\caption{The activity generator.}
		\label{fig: actgen}
		\vspace{-8mm}
	\end{center}
\end{figure}

\textbf{Activity Generator.} Before conducting HLS for an application, the users are required to provide a C-based testbench and a set of input stimuli to verify the correctness of the design output. These files are leveraged in the activity generator. As depicted in Fig.~\ref{fig: actgen}, the activity generator first compiles the given testbench and a library of activity tracking functions written in C++ into object files by the g++ compiler, respectively. In addition, the annotated IR is also converted into an object file by the clang++ compiler. All these object files are further linked together into a single executable file. Through running the executable file with the input vectors, we are able to invoke the target kernel function in the IR, and extract the cycle-level input and output values for each RTL operator into a list. Thereafter, we compute the average switching activity per RTL operator by
\begin{equation}
	\label{eq: sa}
	SA_{op} = \frac{\sum_{i=1}^{M_{op}}{\sum_{j=1}^{N_{op}}{HD(\boldsymbol{s}(i,j),\ \boldsymbol{s}(i,j-1))}}}{M_{op}\cdot N_{op}},
\end{equation}
where $\boldsymbol{s}(i,j)$ is the bit vector for an operand or result $i$ at time step $j$ for the evaluated RTL operator $op$, $M_{op}$ is the total number of operands and results, $N_{op}$ is the length of the list of activity vectors for $op$, and $HD(\cdot)$ is the Hamming distance computation function which counts the differences between two vectors bit by bit.

We further scale the average switching activity for each RTL operator as follows:
\begin{equation}
	\label{eq: nsa}
	SA_{scaled} = \frac{N_{op}}{L} \cdot SA_{op},
\end{equation}
where $L$ is the latency of the target design point estimated by HLS. In this equation, $\frac{N_{op}}{L}$ can be regarded as an activation rate to amortize an operator's average switching activity over the total execution cycles. 

\textbf{Histogram Constructor.} As the directive configurations for different design points lead to different numbers of RTL operators, the size of the currently extracted activity set also varies from design point to design point even for the same application. Noticing that a trained machine learning model is not able to deal with varying feature size, we need to devise a way to convert the set of extracted activities into features so that the feature size is fixed for various design points, and a well developed model is applicable to different applications. To this end, we adopt a \emph{histogram representation} of operator activities. For each opcode, we create a histogram with a pre-defined number of bins, each of which covers a specific activity range. Each RTL operator is first sorted into a particular histogram according to its opcode, and then it is distributed to the bin covering its scaled switching activity, as computed by Eq.~\ref{eq: nsa}. Within each bin, the data statistics to be collected are \emph{the number}, \emph{the percentage} and \emph{the average switching activity} of all the RTL operators in this bin. The fixed-sized statistics for every opcode are used as features and are assembled into an activity feature set for model training and inference. In addition, we adopt the total number of RTL operators for each opcode as a feature.

\subsection{Power Model Generation}
\label{ssec: gpm}
HL-Pow constructs a total number of 256 features, consisting of 11 architecture features mainly accounting for static power and 245 activity features contributing to dynamic power. To obtain ground truth power values for each design point, we conduct RTL implementation flow after the HLS flow, and collect real power measurement during onboard implementation. Besides onboard measurement, gate-level power estimation is another option to get ground truth power values.

We build regression models for power prediction using a variety of supervised learning methods. These models are 1) \emph{linear regression}: classic linear regression and Lasso regression with a $l_1$-norm regularization term; 2) \emph{support vector machine (SVM)}: support vector regression with a radial basis function (RBF) kernel; 3) \emph{tree-based model}: decision tree and ensemble models, including bagging trees, adaboost trees, random forests and gradient boosting decision trees (GBDT); and 4) \emph{neural network}: multi-layer perceptron (MLP), convolutional neural network (CNN) and residual neural network (ResNet). For CNN and ResNet, we construct a 16-by-16 input map from the 256 features, by filling it row by row with architecture features and the total number of RTL operators for each opcode, followed by the other activity features. As for data preprocessing, we perform \emph{data normalization} when necessary. For the first three categories of models, we conduct \emph{feature selection} and \emph{K-fold cross-validation} to determine the models' hyperparameters before model generation. For neural networks, we deploy several widely used model instances, and fine-tune the model hyperparameters.

\section{Algorithm for Design Space Exploration}
\label{sec: dse}
With our power modeling framework, power prediction for a design point can be greatly expedited without running the tedious RTL implementation flow. However, when the goal is to find the Pareto-optimal points from a large design space, there is still a large HLS runtime overhead to exhaustively assess power for every design point through HL-Pow. To tackle this issue, we propose a novel algorithm to approximate the Pareto frontier between latency and power consumption by only sampling a small subset of the design points. Specifically, we apply a priori knowledge generalized from training applications to navigate the search of Pareto-optimal points.

\begin{figure}[t]
	\begin{center}
		\includegraphics[width=0.8\linewidth]{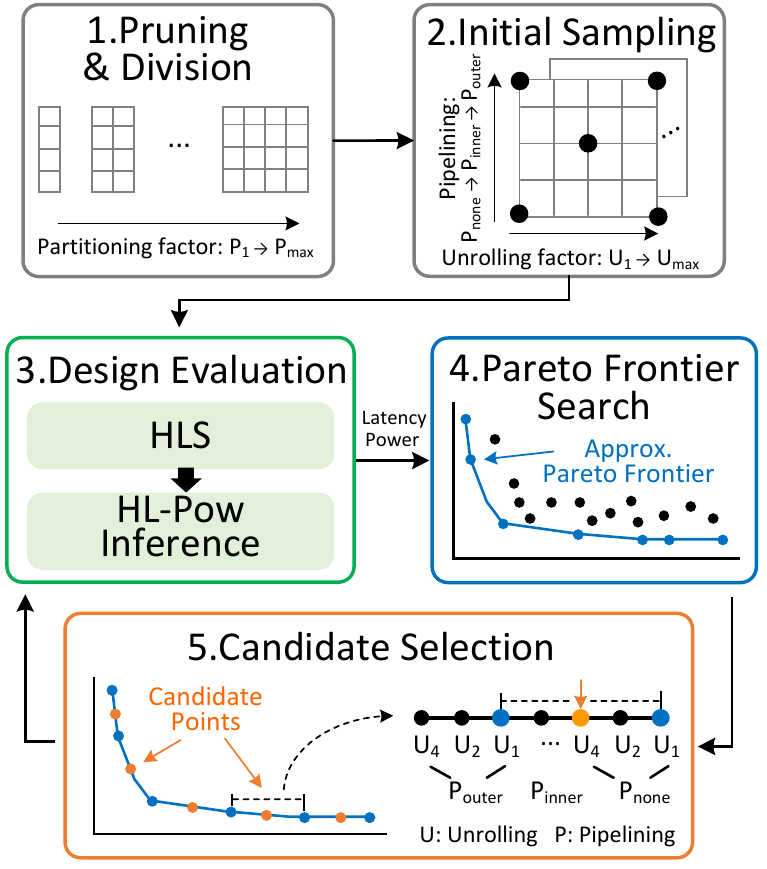}
		\vspace{-2mm}
		\caption{Overveiw of the design space exploration algorithm.}
		\label{fig: dse}
		\vspace{-8mm}
	\end{center}
\end{figure}

The overview of the algorithm is depicted in Fig.~\ref{fig: dse}. We first prune away the design points that produce repetitive RTL designs from the design space, and divide the design space into several regions to explore. The pruning is based on the fact that when an outer loop is pipelined, all  the inner loops are automatically unrolled~\cite{zhao17}. In such a situation, no matter what unrolling factors are set for the inner loops, the resulting architectures are the same as that without unrolling the inner loops. Therefore, we reserve one design point and remove the redundant ones when this situation happens. Afterwards, we split the design space into multiple regions by the array-related directive, namely, array partitioning, and use loop-related directives including loop pipelining and loop unrolling for the search of promising points in each region.

Starting with the trimmed and divided design space, an \emph{initial sampling} step is conducted to collect the first set of design points to assess. The heuristic is to select representative points in each region that are spreading out over the range of both latency and power consumption. Through analysis of the training set, we discover a trend that pipelining the outer loops, compared with pipelininig inner loops or no pipelining, generally leads to higher power consumption along with lower latency. Moreover, unrolling the loops with a larger unrolling factor also brings a similar effect. Following these observations, we can provide a coarse-grained but a priori estimation of latency and power consumption for different directive configurations, and accordingly, we transform each region into a grid-like representation as shown in step 2 of Fig.~\ref{fig: dse}. On top of that, the design points in the corner and in the middle of each grid are selected to add to the initial sampling set, in that they are most likely to demonstrate extreme and median values for both latency and power consumption.

The initial sampling set is fed into HL-Pow to assess latency (by HLS) and power consumption. After obtaining both latency and power values, an approximate Pareto frontier is derived from the current sampling set, and the existing Pareto-optimal points are used as references for identifying promising design points to evaluate. We propose to use the \emph{standard deviation reduction (SDR)}~\cite{quin92} as the metric for candidate point selection. SDR measures the ability that an attribute splits a dataset into subsets: the higher the SDR, the better the dataset is split by similarity. Specific to our case, the dataset is the set of latency or power consumption for all design points in an application, and the attributes are unrolling and pipelining. The SDR in our case can be deduced as
\begin{equation}
	\label{eq: sdr}
	SDR = sd(T) - \sum_i{\frac{|T_i|}{|T|} \times sd(T_i)},
\end{equation}
where $sd(\cdot)$ is the standard deviation computation function, $T$ is the set of latency/power consumption and $T_i$ is the $ith$ subset of $T$ split by unrolling/pipelining. We evaluate all the training applications and find that, for both latency and power consumption, loop pipelining has higher SDR compared to loop unrolling. This means that loop pipelining tends to show a larger effect than loop unrolling on both latency and power consumption, and can better split the design space to indicate differences in both of these metrics. According to this finding, we further transform each region of the design space into an \emph{ordered sequence}, in which the directive configurations are first sorted by loop pipelining in a coarse-grained manner and then by loop unrolling in a fine-grained manner, as shown in step 5 of Fig.~\ref{fig: dse}. In this way, the latency/power consumption can be roughly estimated as monotonic decreasing/increasing following the direction from right to left in this representation. 

We identify each pair of neighboring points in the approximate Pareto set that are from the same region, and annotate them in the corresponding ordered sequence. For each pair of annotated points, we locate the middle point between them in the sequence and add it to the sampling set. If this middle point has already been added to the sampling set, we remove it from the sequence, and instead search for the updated middle point to add. The above steps, namely, \emph{design evaluation}, \emph{Pareto frontier search} and \emph{candidate selection}, are iterated to search for promising design points until a user-defined budget of HLS runs is reached or no more candidates exist. Finally, to ensure that the real Pareto-optimal points are not pruned away due to the error induced by power estimation, we allow the design points within a pre-defined deviation (e.g., 5\%) of the power consumption from the nearest Pareto-optimal points to be incorporated into the Pareto set.

\section{Experimental Results}
\subsection{Experimental Setup}
The HL-Pow design flow is fully automated and implemented with Python and C++ for feature construction, model establishment and power inference. Different types of learning models are realized in Scikit-learn~\cite{scikit}, XGBoost~\cite{xgboost} and Keras~\cite{keras}, respectively. We apply our design flow to evaluate 22 applications from different categories in Polybench~\cite{polybench}, resulting in up to 11326 valid design points and 256 features per design point. The design points are synthesized using floating-point arithmetic and implemented under a timing constraint of 10 ns. The FPGA  development toolkit we use is Xilinx Vivado Design Suite 2018.2. We implement all the design points on a Xilinx Ultrascale+ ZCU102 FPGA board and collect real power consumption through onboard measurement with the Power Advantage Tool~\cite{powtool}. We customize the HLS optimization strategies that fit the applications into the target platform, as shown in Table~\ref{table: diropt}.

\begin{table}[t]
	\centering
	\caption{Directive options suitable for the target platform.}
	\vspace{-2mm}
	\label{table: diropt}
	\begin{tabular}[width=\linewidth]{c|c}
		\toprule
		\multicolumn{1}{c|}{\textbf{Directive}} & \multicolumn{1}{c}{\textbf{Option}}\\
		\midrule
		\multicolumn{1}{c|}{Array partitioning} & type: cyclic; factor: [1, 2, 4, 8]\\
		\multicolumn{1}{c|}{Loop pipelining} & different levels of nested loops\\
		\multicolumn{1}{c|}{Loop unrolling} & factor: [1, 2, 4, 8]\\
		\bottomrule
	\end{tabular}
	\vspace{-2mm}
\end{table}

\subsection{Performance of Power Modeling}
We use 8784 design points from 15 applications for training and validation, and 2542 design points (\textgreater 20\%) from the other seven applications (Atax, Bicg, Fdtd\_2d, Gemm, Gramschmidt, Jacobi\_2d, Mvt) are used for testing only. The applications for testing are from different categories of Polybench and are unseen in the training set. This ensures that the machine learning models we build are not specifically tuned for the test cases. We evaluate the four categories of machine learning models discussed in Section~\ref{ssec: gpm}, and show the results of the best model from each category in Table~\ref{table: acc}. The model performance is measured by mean absolute error (MAE) in percentage terms. The CNN (based on Keras CIFAR-10 CNN) achieves the best overall performance among all the learning models, leading to a prediction error of 4.67\% (24.03 mW). The GBDT also demonstrates good performance that is comparable to the CNN. For Atax, Bicg, Gemm and Mvt applications, the GBDT even outperforms the CNN. In contrast, the Lasso linear regression and SVM give rise to much higher error in power modeling. This conforms to the conclusion in prior studies~\cite{bogl00, lee15} that the power behavior of complex hardware designs is generally non-linear. Compared with the closest state-of-the-art work~\cite{zuo15} that incurs a 5.04\% error, HL-Pow is more generic, accurate and user-friendly by adopting a one-time modeling process, obviating the need for iterative RTL-based power characterization per target design.

\begin{table}[t]
	\centering
	\vspace{-1mm}
	\caption{Accuracy of power modeling.}
	\vspace{-2mm}
	\label{table: acc}
	\begin{tabular}[width=\linewidth]{c|c|c|c|c|c}
		\toprule
		\multirow{2}{*}{\textbf{Application}} & \multirow{1}{*}{\textbf{Power}} & \multicolumn{4}{c}{\textbf{MAE (\%) of Learning Models}}\\
		& \textbf{Range (W)} & \multicolumn{1}{c|}{Lasso} & \multicolumn{1}{c|}{SVM} & \multicolumn{1}{c|}{GBDT} & \multicolumn{1}{c}{CNN}\\
		\midrule
		\multicolumn{1}{c|}{Atax} & 0.30 -- 1.00 & 7.46 & 15.07 & 2.80 & 5.14\\
		\multicolumn{1}{c|}{Bicg} & 0.30 -- 1.15 & 6.21 & 20.62 & 4.63 & 7.80\\
		\multicolumn{1}{c|}{Fdtd\_2d} & 0.29 -- 1.36 & 9.46 & 10.81 & 4.79 & 3.98\\
		\multicolumn{1}{c|}{Gemm} & 0.30 -- 0.86 & 6.92 & 17.51 & 3.69 & 5.15\\
		\multicolumn{1}{c|}{Gramschmidt} & 0.29 -- 0.65 & 9.07 & 12.31 & 6.26 & 5.69\\
		\multicolumn{1}{c|}{Jacobi\_2d} & 0.30 -- 1.31 & 10.67 & 14.16 & 6.32 & 4.36\\
		\multicolumn{1}{c|}{Mvt} & 0.30 -- 1.09 & 9.58 & 14.03 & 4.11 & 4.40\\
		\midrule
		\multicolumn{1}{c|}{Overall} & 0.29 -- 1.36 & 9.08 & 13.00 & 4.78 & 4.67\\
		\bottomrule
	\end{tabular}
	\vspace{-5mm}
\end{table}

\vspace{-1mm}
\subsection{Quality of Design Space Exploration}
We investigate the quality of our DSE algorithm, as proposed in Section~\ref{sec: dse}, with the three applications from the test set (Fdtd\_2d, Mvt and Gramschmidt) that have the largest number of design points. To assess the performance of our DSE algorithm in real cases, we calibrate the Pareto-optimal points in the approximate Pareto set using the corresponding real power values from measurement. \emph{Average distance from reference set (ADRS)} is used as the metric to quantify the difference between the approximate and the exact Pareto sets. ADRS is defined as
\begin{equation}
	\small
	\begin{split}
		\label{eq: adrs}
		& ADRS(\bar{P}, P) = \left[ \frac{1}{|P|}\sum_{p \in P}\min_{\bar{p} \in \bar{P}}(\delta(\bar{p}, p)) \right] \times 100\%, \\
		& \delta(\bar{p}, p) = \max \left\{0, \frac{Lat_{\bar{p}}-Lat_p}{Lat_p}, \frac{Pwr_{\bar{p}}-Pwr_p}{Pwr_p}\right\},
	\end{split}
\end{equation}
where $\bar{P}$ is the approximate Pareto set, $P$ is the exact Pareto set, and $Lat$ and $Pwr$ denote latency and power, respectively. The lower the ADRS, the smaller the difference between the approximate Pareto set and the exact Pareto set. 

\begin{figure}[t]
	\begin{center}
		\includegraphics[width=0.4938\linewidth]{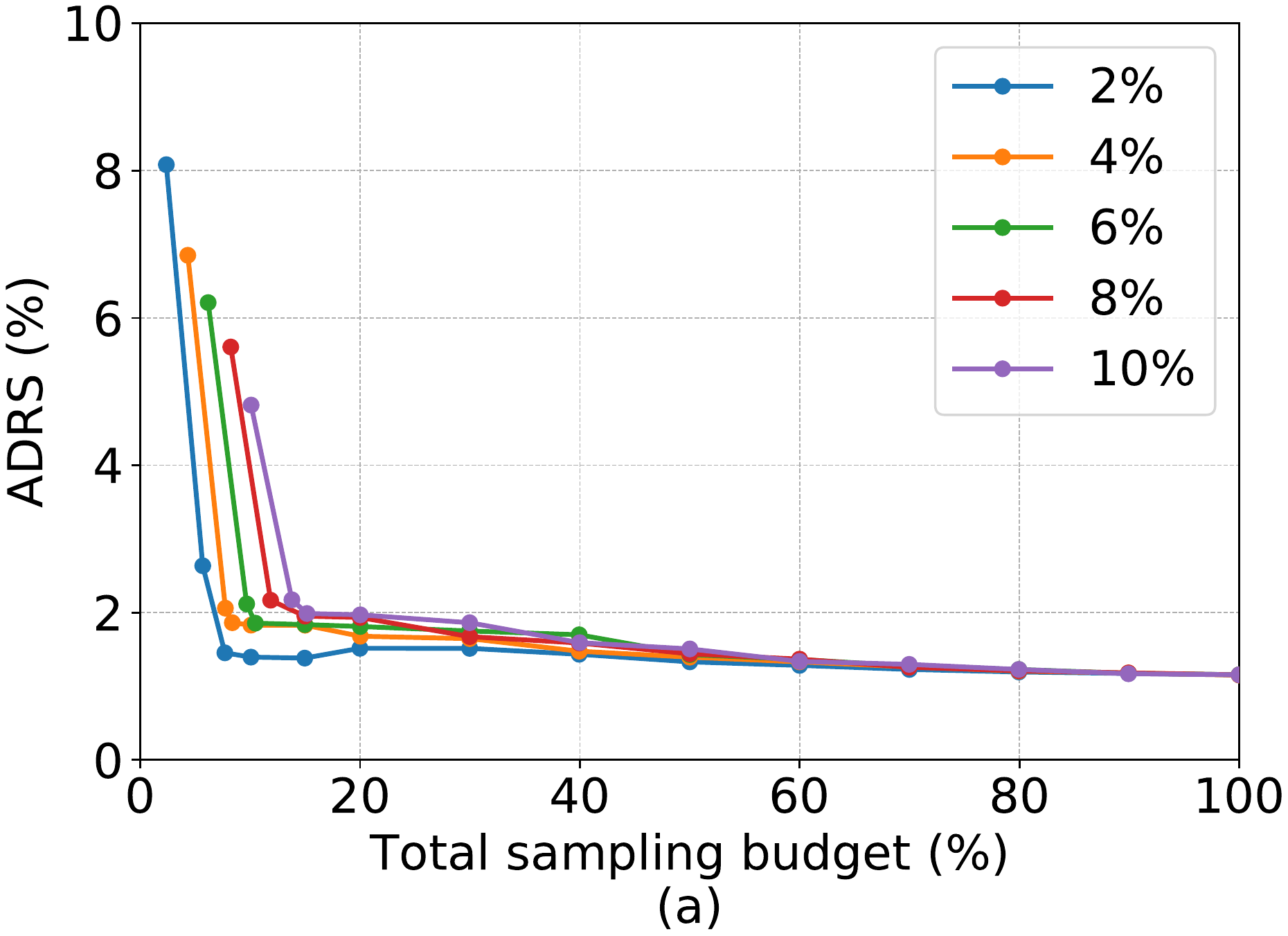} 
		\includegraphics[width=0.4938\linewidth]{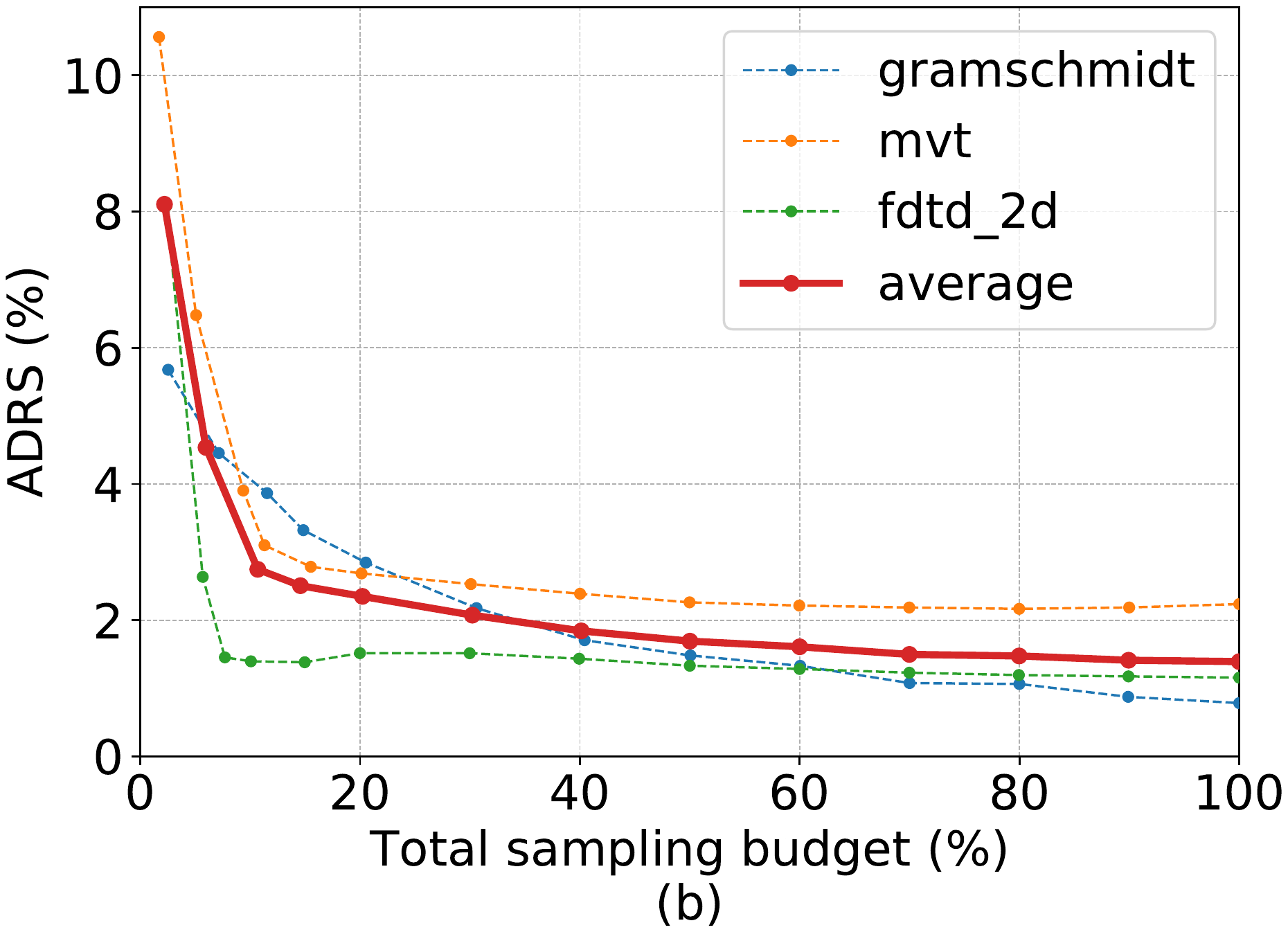}\\
		\includegraphics[width=0.4934\linewidth]{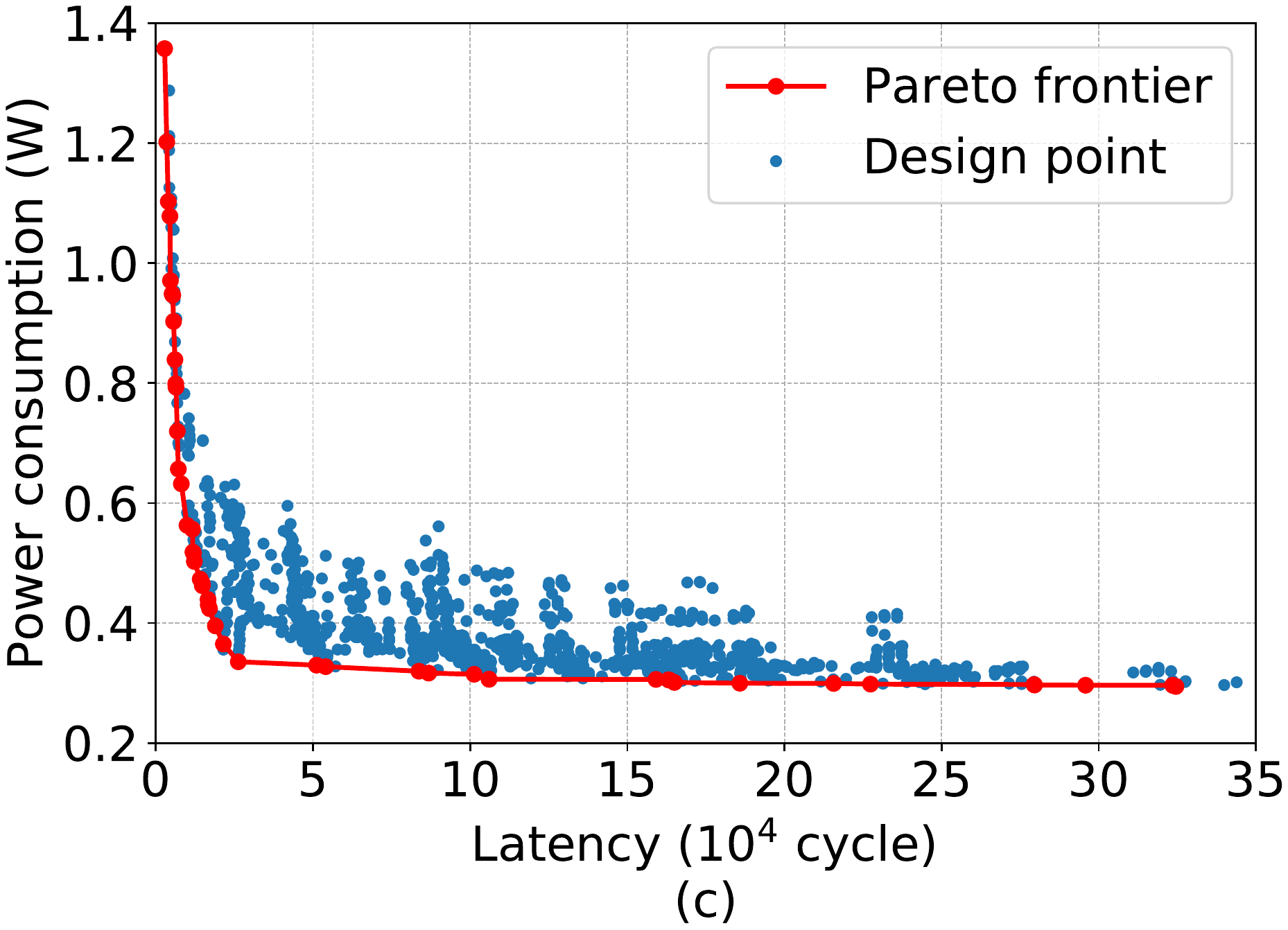} 
		\includegraphics[width=0.4934\linewidth]{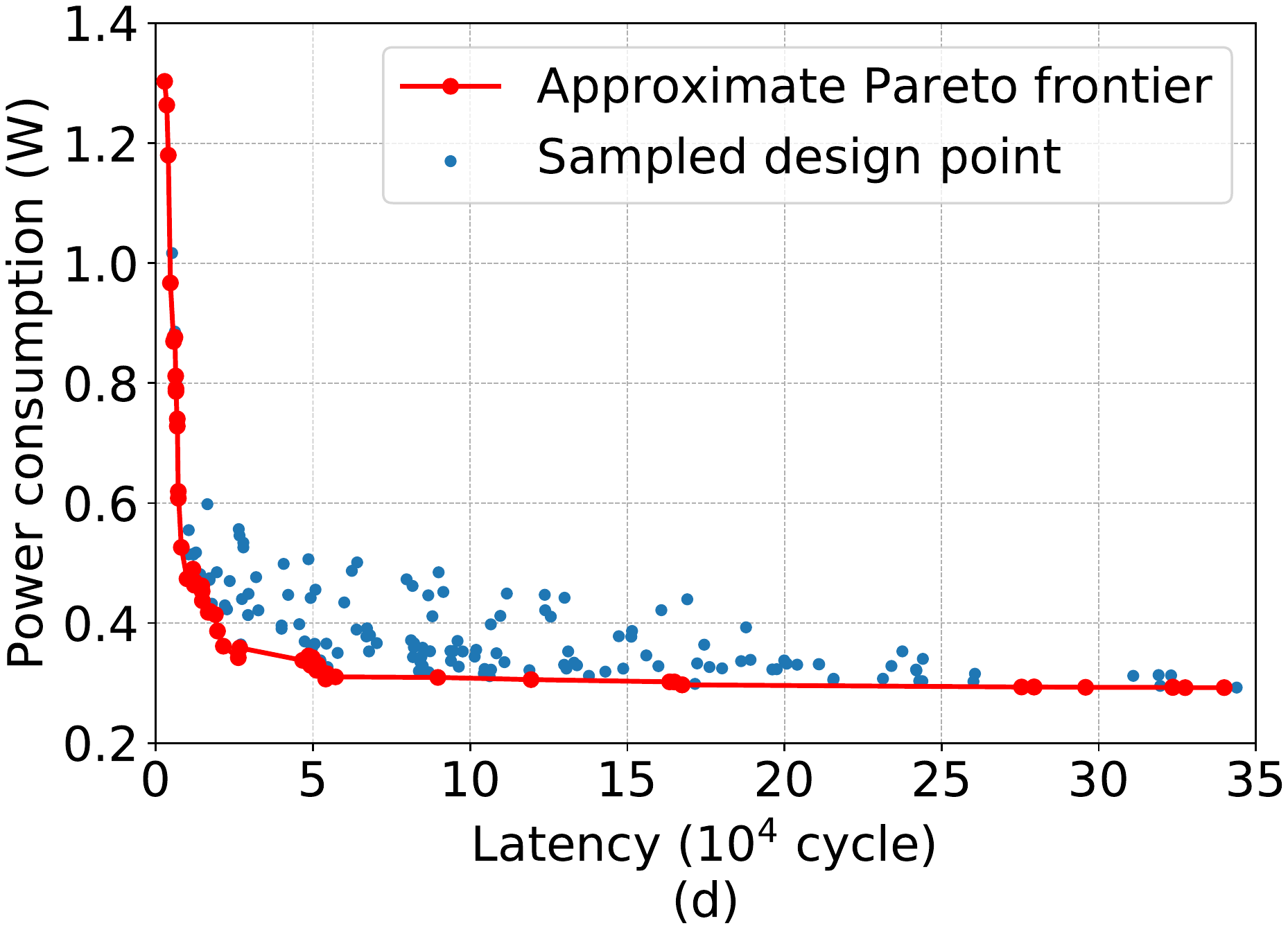}
		\vspace{-8mm}
		\caption{Results of Pareto frontier approximation: (a) ADRS of Fdtd\_2d application with different initial sampling rate; (b) ADRS of different sampling budgets under a 2\% initial sampling rate; (c) real Pareto frontier of Fdtd\_2d with the complete sample set; and (d) approximate Pareto frontier of Fdtd\_2d with a 2\% initial sampling rate and 20\% sampling budget. }
		\label{fig: pareto}
	\end{center}
	\vspace{-8mm}
\end{figure}

We investigate how different initial sampling rates and total sampling budgets (i.e., the proportion of design points for sampling) affect the quality of approximation results. We first evaluate the initial sampling rates from 2\% to 10\%. Fig.~\ref{fig: pareto} (a) depicts the results for the application with the largest number of design points, Fdtd\_2d, and the other applications indicate a similar trend. ADRS decreases rapidly as the total sampling budget increases from a small starting point, which showcases the efficacy of our DSE algorithm. Moreover, we can observe that applying different initial sampling rates leads to a converged ADRS as the sampling budget increases. Nevertheless, using a small initial sampling rate benefits the approximation quality given a limited sampling budget. This is because it effectively balances the sampling proportion between initial sampling and iterative searching. As a result, we adopt a 2\% initial sampling rate in the following experiments. 

The ADRS for different applications is shown in Fig.~\ref{fig: pareto} (b). Our algorithm demonstrates good results with a sampling budget of 20\% and converges at a sampling budget of 40\%, resulting in an average ADRS of 2.35\% and 1.84\%, respectively. Fig.~\ref{fig: pareto} (c) and (d) show the real and approximate Pareto frontiers for Fdtd\_2d, respectively. From them, we can observe a clear tradeoff between latency and power consumption. Fig.~\ref{fig: pareto} (d) also indicates good approximation quality. In brief, our DSE algorithm can approach a close approximation of the real Pareto frontier with a small sampling budget.

\section{Conclusion}
Power consumption is a key consideration for hardware designs. However, existing methodologies for power estimation or measurement incur high development cost and also exhibit many restrictions. In light of these problems, we target efficient and accurate power estimation for FPGA designs at an early design stage. We introduce HL-Pow, a learning-based power modeling framework for HLS. We first propose an automated and fast feature construction flow to capture informative features for power indication, simply based upon HLS results, and then present a modeling framework which can build a generic power model that works for diverse designs without the necessity of model regeneration. HL-Pow can significantly accelerate the power prediction process for FPGA designs as the execution of the time-consuming RTL implementation flow can be skipped. Experimental results verify that HL-Pow can achieve an average prediction error within 4.67\% of onboard power measurement. Based on HL-Pow, we describe a novel and efficient algorithm to explore the tradeoff between latency and power consumption of HLS designs. The proposed algorithm retrieves a close approximation of the real Pareto frontier with an average ADRS of 2.35\% and 1.84\% while only sampling 20\% and 40\% of design points, respectively, in the complete design space. 

\section*{Acknowledgment}
This work is funded by Hong Kong RGC GRF under grant 16245116.

\bibliographystyle{IEEEtran}
\bibliography{ref}

\begin{thebibliography}{10}
\providecommand{\url}[1]{#1}
\csname url@samestyle\endcsname
\providecommand{\newblock}{\relax}
\providecommand{\bibinfo}[2]{#2}
\providecommand{\BIBentrySTDinterwordspacing}{\spaceskip=0pt\relax}
\providecommand{\BIBentryALTinterwordstretchfactor}{4}
\providecommand{\BIBentryALTinterwordspacing}{\spaceskip=\fontdimen2\font plus
\BIBentryALTinterwordstretchfactor\fontdimen3\font minus
  \fontdimen4\font\relax}
\providecommand{\BIBforeignlanguage}[2]{{%
\expandafter\ifx\csname l@#1\endcsname\relax
\typeout{** WARNING: IEEEtran.bst: No hyphenation pattern has been}%
\typeout{** loaded for the language `#1'. Using the pattern for}%
\typeout{** the default language instead.}%
\else
\language=\csname l@#1\endcsname
\fi
#2}}
\providecommand{\BIBdecl}{\relax}
\BIBdecl

\bibitem{cous09}
P.~{Coussy} \emph{et~al.}, ``An introduction to high-level synthesis,''
  \emph{IEEE Design Test of Computers}, pp. 8--17, 2009.

\bibitem{liu13}
H.-Y. Liu and L.~P. Carloni, ``On learning-based methods for design-space
  exploration with high-level synthesis,'' in \emph{Proc. of DAC}, 2013.

\bibitem{fer18}
L.~{Ferretti} \emph{et~al.}, ``Lattice-traversing design space exploration for
  high level synthesis,'' in \emph{Proc. of ICCD}, 2018.

\bibitem{gzhong17}
G.~{Zhong} \emph{et~al.}, ``Design space exploration of {FPGA-based}
  accelerators with multi-level parallelism,'' in \emph{Proc. of DATE}, 2017.

\bibitem{meng16}
P.~Meng \emph{et~al.}, ``Adaptive threshold non-pareto elimination: Re-thinking
  machine learning for system level design space exploration on {FPGAs},'' in
  \emph{Proc. of DATE}, 2016.

\bibitem{dong16}
{Dong Liu} and B.~C. {Schafer}, ``Efficient and reliable high-level synthesis
  design space explorer for {FPGAs},'' in \emph{Proc. of FPL}, 2016.

\bibitem{lfer18}
L.~{Ferretti} \emph{et~al.}, ``Cluster-based heuristic for high level synthesis
  design space exploration,'' \emph{IEEE Transactions on Emerging Topics in
  Computing}, 2018.

\bibitem{vivadohls}
{Xilinx Ltd}, ``{Vivado design suite user guide: High level synthesis},''
  \emph{Xilinx White Paper}, April 2017.

\bibitem{liu94}
D.~Liu and C.~Svensson, ``Power consumption estimation in {CMOS VLSI} chips,''
  \emph{IEEE Journal of Solid-State Circuits}, 1994.

\bibitem{lee15}
D.~Lee \emph{et~al.}, ``Learning-based power modeling of system-level black-box
  {IPs},'' in \emph{Proc. of ICCAD}, 2015, pp. 847--853.

\bibitem{zhe18}
Z.~{Lin} \emph{et~al.}, ``An ensemble learning approach for in-situ monitoring
  of {FPGA} dynamic power,'' \emph{TCAD}, 2018.

\bibitem{zuo15}
W.~Zuo \emph{et~al.}, ``A polyhedral-based {SystemC} modeling and generation
  framework for effective low-power design space exploration,'' in \emph{Proc.
  of ICCAD}, 2015.

\bibitem{liang18}
Y.~{Liang} \emph{et~al.}, ``{FlexCL}: A model of performance and power for
  {OpenCL} workloads on {FPGAs},'' \emph{TC}, 2018.

\bibitem{bogl00}
A.~Bogliolo \emph{et~al.}, ``Regression-based {RTL} power modeling,''
  \emph{TODAES}, 2000.

\bibitem{shao14}
Y.~S. Shao \emph{et~al.}, ``Aladdin: A pre-{RTL}, power-performance accelerator
  simulator enabling large design space exploration of customized
  architectures,'' in \emph{Proc. of ISCA}, 2014.

\bibitem{dm07}
D.~Chen \emph{et~al.}, ``High-level power estimation and low-power design space
  exploration for {FPGAs},'' in \emph{Proc. of ASP-DAC}, 2007.

\bibitem{hao15}
H.~{Liang} \emph{et~al.}, ``Hierarchical library based power estimator for
  versatile {FPGAs},'' in \emph{Proc. of IEEE International Symposium on
  Embedded Multicore/Many-core Systems-on-Chip}, 2015.

\bibitem{llvm04}
C.~Lattner and V.~Adve, ``{LLVM}: A compilation framework for lifelong program
  analysis \& transformation,'' in \emph{Proc. of CGO}, 2004.

\bibitem{zhao17}
J.~Zhao \emph{et~al.}, ``{COMBA}: A comprehensive model-based analysis
  framework for high level synthesis of real applications,'' in \emph{Proc. of
  ICCAD}, 2017.

\bibitem{quin92}
J.~R. Quinlan \emph{et~al.}, ``Learning with continuous classes,'' in
  \emph{Proc. of Australian Joint Conference on Artificial Intelligence}, 1992.

\bibitem{scikit}
F.~Pedregosa \emph{et~al.}, ``{Scikit-learn: Machine learning in Python},''
  \emph{J. Mach. Learning Research}, 2011.

\bibitem{xgboost}
T.~Chen and C.~Guestrin, ``{XGBoost}: A scalable tree boosting system,'' in
  \emph{Proc. of KDD}, 2016.

\bibitem{keras}
\BIBentryALTinterwordspacing
F.~Chollet \emph{et~al.} (2018) Keras: The python deep learning library.
  [Online]. Available: \url{https://keras.io/}
\BIBentrySTDinterwordspacing

\bibitem{polybench}
\BIBentryALTinterwordspacing
L.-N. Pouchet. (2012) Polybench: The polyhedral benchmark suite. [Online].
  Available: \url{http://www.cs.ucla.edu/pouchet/software/polybench}
\BIBentrySTDinterwordspacing

\bibitem{powtool}
{Xilinx Ltd}, ``Zynq {UltraScale+ MPSoC Power Advantage Tool} 2018.1,''
  \emph{Xilinx Wiki}, 2018.

\end{thebibliography}

\end{document}